\documentclass[12pt,A4paper]{article}

\usepackage{epsfig,amssymb}

\begin{document}

\title{\bf Memory   in the aging of a polymer glass}
\author{L. Bellon,  S. Ciliberto and  C. Laroche \\
         Ecole Normale
Sup\'erieure de Lyon, Laboratoire de Physique ,\\
 C.N.R.S. UMR5672,  \\ 46, All\'ee d'Italie, 69364 Lyon Cedex
07,  France\\
        }
 \maketitle

\begin{abstract}
Low frequency  dielectric measurements on plexiglass (PMMA) show
that cooling and heating the sample at constant rate give an
hysteretic dependence on temperature of the dielectric constant
$\epsilon$.  A temporary stop of cooling produces a downward
relaxation of $\epsilon$. Two main features are observed i) when
cooling is resumed $\epsilon$ goes back to the values  obtained
without the cooling stop ( i.e. the low temperature state is
independent of the cooling history) ii) upon reheating $\epsilon$
keeps the memory of the aging history ({\it Memory}). The
analogies and differences with similar experiments done in spin
glasses are discussed.
\end{abstract}

\medskip
{\bf PACS:} 75.10.Nr, 77.22Gm, 64.70Pf, 05.20$-$y.


\newpage

The aging of glassy materials is a widely studied phenomenon
\cite{Struick,book}, which is characterized by a slow evolution of
the system toward equilibrium, after a quench below the glass
transition temperature $T_g$. In other words the properties of
glassy materials depend on the time spent at a temperature smaller
than $T_g$.  In spite of the interesting experimental
\cite{Vincent,Alberici,Nagel,VincentPRL} and theoretical progress
\cite{book,Bouchaud,Mezard}, done in the last years, the physical
mechanisms of aging  are not yet fully understood. In fact on  the
basis of  available experimental data it is very difficult to
distinguish which  is the most suitable theoretical approach for
describing  the aging processes of different materials. In order
to give more insight into this problem several experimental
procedures have been proposed and applied to the study of the
aging of various materials, such as spin-glasses
(SG)\cite{Vincent,VincentPRL,Jonsson1}, orientational glasses
(OG)\cite{Alberici,Doussineau}, polymers \cite{Struick,Bellon} and
supercooled liquids (SL)\cite{Nagel}. Among these procedures we
may recall the applications of small temperature cycles to a
sample during the aging time\cite{Vincent,Alberici,Nagel,Bellon}.
These experiments have shown three main results in different
materials: i)  there is an important difference between positive
and negative cycles and  the details of the response to these
perturbations are material dependent
\cite{Vincent,Alberici,Bellon}; ii) for SG \cite{Vincent} the time
spent at the higher temperature does not contribute to the aging
at a lower temperature whereas for plexiglass (PMMA) \cite{Bellon}
and OG \cite{Alberici} it slightly modifies the long time
behavior; iii) A memory effect has been observed for negative
cycles. Specifically when temperature goes back to the high
temperature the system recovers its state before perturbation. In
other words the time spent at low temperature does not contribute
to the aging behavior at the higher temperature.

These results clearly exclude models based on the activation
processes over temperature  independent barriers, where the time
spent at high temperature would help to find easily the
equilibrium state. At the same time it is difficult to decide
which is the most appropriate theoretical approach to describe the
response to these temperature cycles \cite{Vincent}. For example a
recent model explains the results in SG but not in OG and in  PMMA
\cite{Kurchan}. In order to have a better understanding of the
free energy landscape of SG and OG, a  new cooling protocol has
been proposed  \cite{VincentPRL} and  used in several experiments
\cite{VincentPRL,Jonsson1,Doussineau}. This protocol, which is
characterized by a temporary cooling stop, has revealed that in SG
and in  OG  the low temperature state is independent of the
cooling history \footnote{This effect was named {\it Chaos} in
ref.\cite{VincentPRL} but in a more recent paper \cite{VincentEPL}
the relevance  of chaos for  aging has been disputed. The idea of
chaos  in this context might be inappropriate}
 and that these materials keep the memory of the aging
history ({\it Memory effect}) \cite{VincentPRL}.

  The purpose of this
letter is to describe an experiment where we use the  cooling
protocol, proposed in ref.\cite{VincentPRL}, to show that a
 memory  effect is  present  during the aging of the
dielectric constant of plexiglass (PMMA), which is a polymer glass
with $T_g=388K$
   \cite{bookp,comment}.  We also
compare the behavior of PMMA to that of SG and OG, submitted to
the same cooling protocol.

To determine the dielectric constant, we measure the complex
impedance of a capacitor whose dielectric is the PMMA sample. In
our experiment a disk of PMMA of diameter $10cm$ and thickness
$0.3mm$ is inserted between the plates of a capacitor whose vacuum
capacitance is $C_o=230 pF$.The capacitor temperature is stable
within $0.1 K$  and it may be changed  from $300 K$ to $500K$.

The capacitor is a component of the feedback loop of a precision
voltage amplifier whose input is connected to a signal generator.
We obtain the real and imaginary part of the capacitor impedance
by measuring the response of the amplifier to a sinusoidal  input
signal. This apparatus allows us to measure the real and imaginary
part of the dielectric constant $\epsilon=\epsilon_1 +i \
\epsilon_2$ as a function of temperature $T$,  frequency $\nu$ and
time $t$. Relative variations  of $\epsilon$ smaller than  $
10^{-3}$ can be measured in all the frequency range used in this
experiment, i.e. $0.1Hz < \nu <100Hz$.  The following discussion
will focus only on $\epsilon_1$ , because the behavior of
$\epsilon_2$ leads to the same conclusions.

 The measurement is
performed in the following way. We first reinitialize the PMMA
history by heating the sample at a temperature $T_{max}>T_g$. The
sample is left at $T_{max}=415K$ for a few hours. Then it is
slowly cooled from $T_{max}$ to a temperature $T_{min}=313K$ at
the constant  rate $|R|=|{\partial T\over
\partial t}|$ and heated back to $T_{max}$ at the same
 $|R|$. The dependence of $\epsilon_1$ on $T$
 obtained  by cooling and heating the sample at a constant
$|R|$, is called the reference curve $\epsilon_r$.

\begin{figure}
\begin{center}
\end{center}
\centerline{ \epsfig{figure=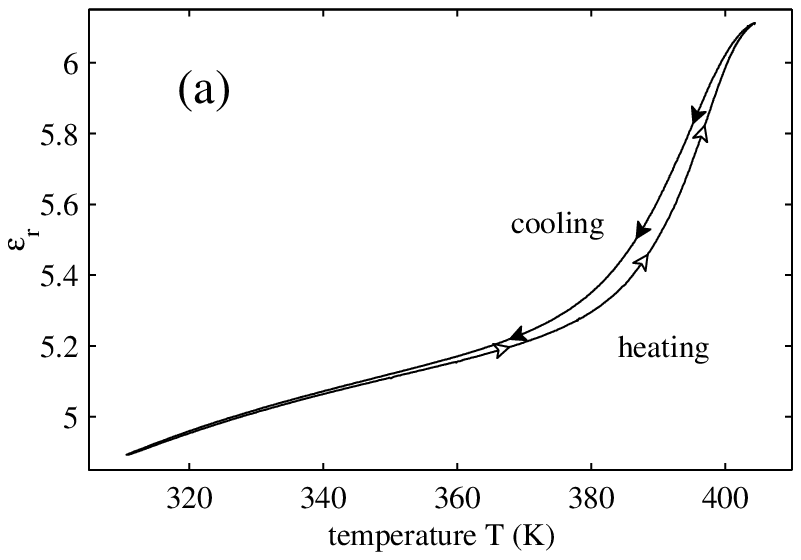}
\epsfig{figure=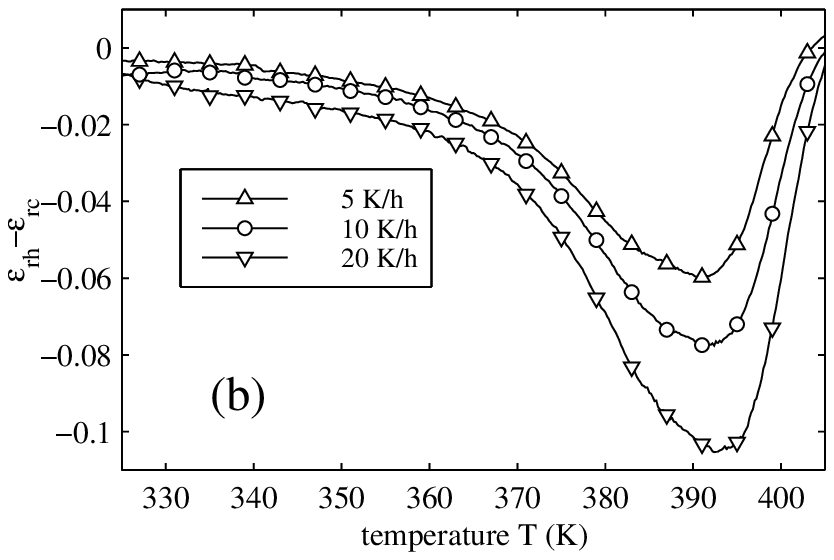}}

\caption{(a) Evolution of $\epsilon_r$ at $\nu=0.1Hz$ as a
function of $T$. Reference curve obtained with $|R|=20K/h$. (b)
Hysteresis of the reference curve (difference between the heating
and cooling curves $\epsilon_{rh} - \epsilon_{rc}$) for 3
different $|R|$ : $5 K/h$ ($\vartriangle$), $10 K/h$ ($\circ$) and
$20 K/h$ ($\triangledown$).} \label{fig:tempcyc1}
\end{figure}

 As  an example of reference curve
  we plot in fig.\ref{fig:tempcyc1}(a) $\epsilon_r$, measured at
  $0.1Hz$ and at $|R|=20K/h$.  We see that $\epsilon_r$
 presents a
hysteresis between the cooling and the heating in the interval
$350K<T<405K$. This hysteresis depends on the cooling and heating
rates. Indeed, in fig.\ref{fig:tempcyc1}(b), the difference
between the heating curve ($\epsilon_{rh}$) and the cooling curve
($\epsilon_{rc}$) is plotted as a function of $T$ for different
$|R|$. The faster we change temperature, the bigger hysteresis we
get. Furthermore the temperature  of the hysteresis maximum is a
few degrees above $T_g$, specifically at $T\approx392K$. The
temperature of this maximum gets closer to $T_g$ when the rate is
decreased.

\begin{figure}
\begin{center}
\centerline{\epsfig{figure=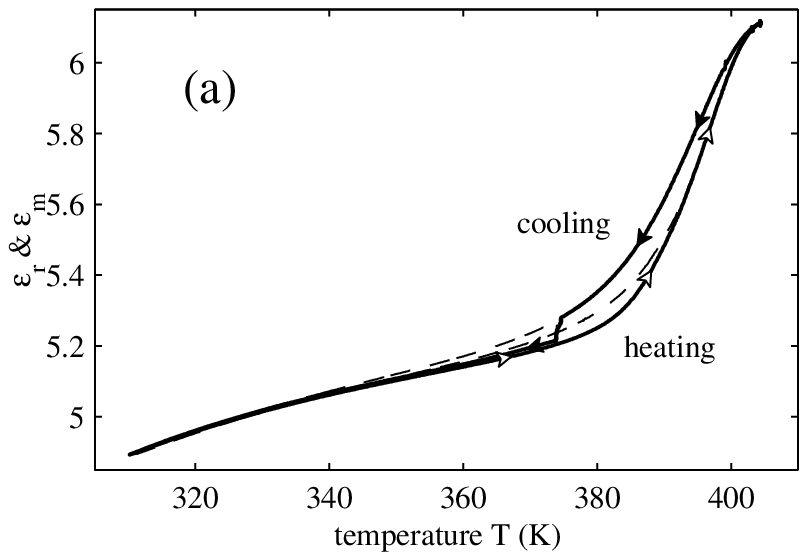}
\epsfig{figure=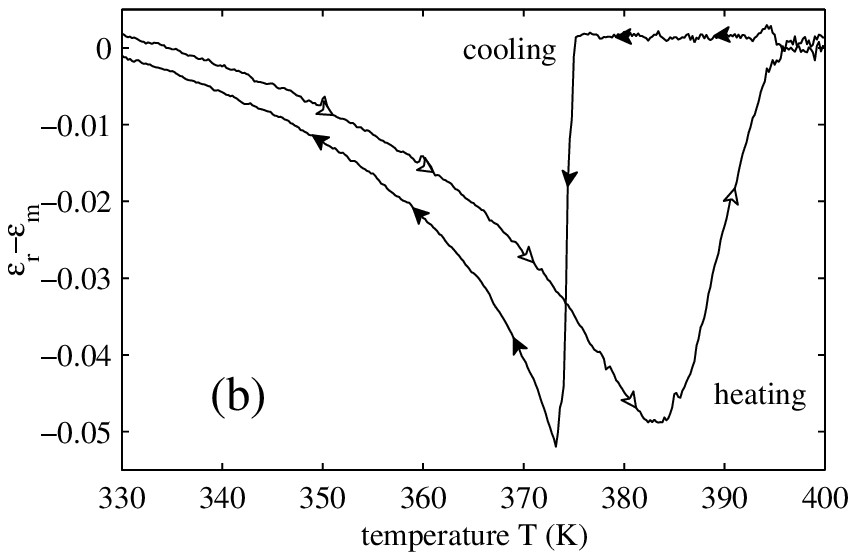}}
\end{center}

\caption{(a)Evolution of $\epsilon$ at $\nu=0.1Hz$ as a function
of $T$. The dashed line corresponds to the reference curve
($\epsilon_r$) of Fig.\ref{fig:tempcyc1}(a). The solid bold line
corresponds to a different cooling procedure : the sample is
cooled, at $R=-20K/h$, from $T_{max}$ to $T_{stop}=374K$, where
cooling is  stopped for $10h$. Afterwards the sample is  cooled at
the same $R$ till $T_{min}$ and then heated again at $R=20K/h$
till $T_{max}$. (b) Difference between the evolution of
$\epsilon_r$ and  $\epsilon_m$. Downward filled arrows correspond
to cooling ($\epsilon_{mc}-\epsilon_{rc}$) and upward empty arrows
to heating ($\epsilon_{mh}-\epsilon_{rh}$).} \label{fig:tempcyc2}
\end{figure}

We neglect for the moment the rate dependence of the  hysteresis
and we consider as reference curve the one, plotted in
fig.\ref{fig:tempcyc1}(a), which has been obtained at $\nu=0.1Hz$
and at $|R|=20K/h$.
 The evolution of $\epsilon_1$ can be quite different from
$\epsilon_r$ if we use  the temperature cycle proposed in
ref.\cite{VincentPRL}. After a cooling at $R=-20K/h$ from
$T_{max}$ to $T_{stop}=374K$ the sample is maintained at
$T_{stop}$ for $10h$. After this time interval the sample is
cooled again, at the same $R$, down to $T_{min}$. Once the sample
temperature reaches $T_{min}$ the sample is heated again at
$R=20K/h$ up to $T_{max}$. The dependence of $\epsilon_1$ as a
function of $T$, obtained when the sample is submitted to this
temperature cycle with the cooling stop at $T_{stop}$, is called
the memory curve $\epsilon_m$. In fig.\ref{fig:tempcyc2}(a),
$\epsilon_m$ (solid line), measured at $\nu=0.1 Hz$, is plotted as
a function of $T$. The dashed line corresponds to the reference
curve of fig.\ref{fig:tempcyc1}(a).
 We notice that
$\epsilon_m$ relaxes downwards when cooling is stopped at
$T_{stop}$: this corresponds to the vertical line  in
fig.\ref{fig:tempcyc2}(a) where $\epsilon_m$ departs from
$\epsilon_r$. When cooling is resumed $\epsilon_1$ merges into
$\epsilon_r$ for $T <340K$. The aging at $T_{stop}$ has not
influenced the result at low temperature.

\begin{figure}
\begin{center}
\centerline{\epsfig{figure=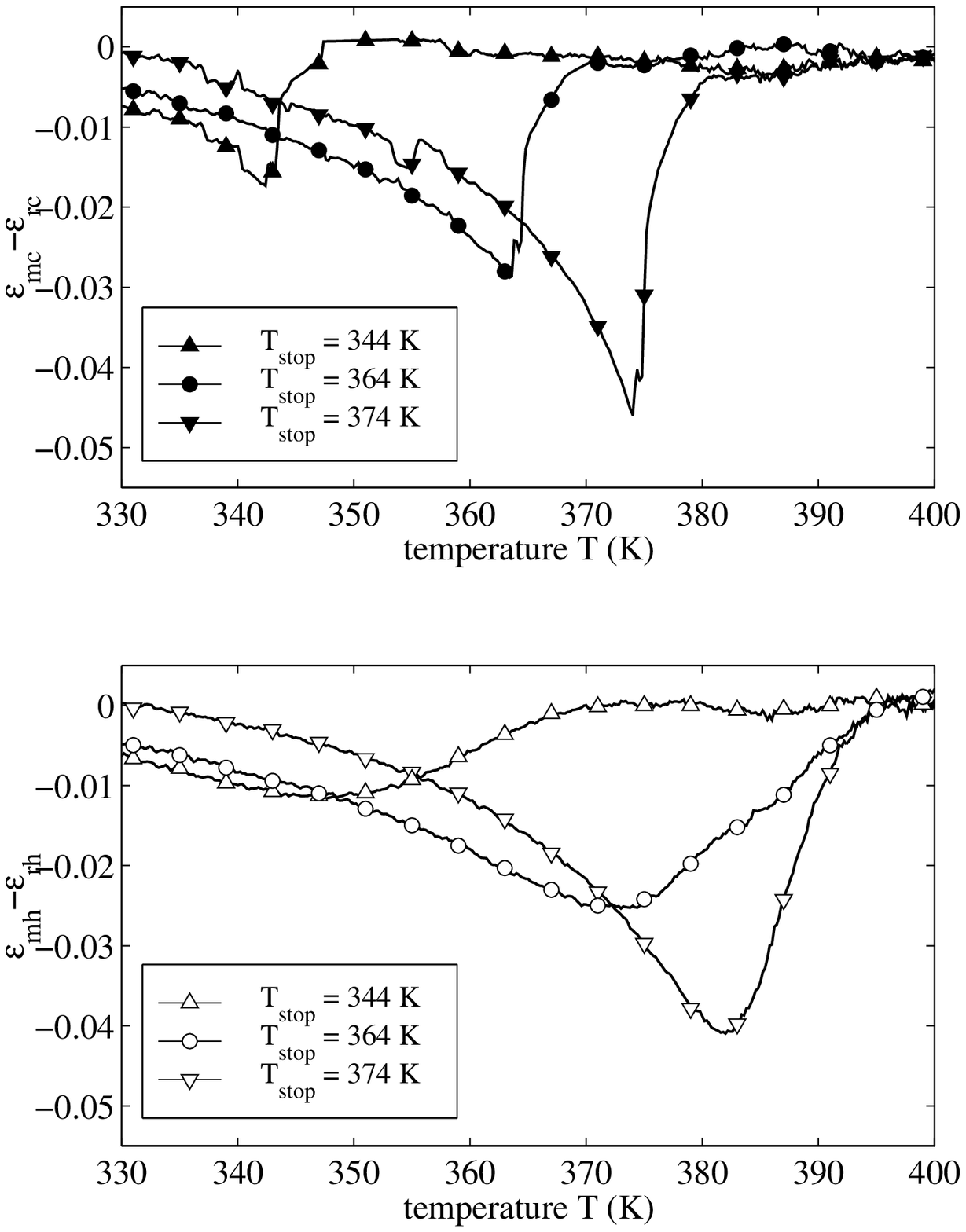}}
\end{center}

\caption{Dependence on $T_{stop}$. Difference between $\epsilon_r$
and $\epsilon_m$ for three different cooling stops at $\nu=0.1Hz$,
$|R|=10K/h$ and $t_{stop}=10h$.  (a) Writing memory (cooling) :
$\epsilon_{mc} - \epsilon_{rc}$ with $T_{stop}=344 K$
($\blacktriangle$), $T_{stop}=364 K$ ($\bullet$) and
$T_{stop}=374K$ ($\blacktriangledown$). (b) Reading memory
(heating) : $\epsilon_{mh} - \epsilon_{rh}$ of  $T_{stop}=344 K$
($\vartriangle$), $T_{stop}=364 K$ ($\circ$) and$T_{stop}=374 K$
($\triangledown$).
 }
\label{fig:Tstop}
\end{figure}

 During the heating period  the system
keeps the memory of the aging  at $T_{stop}$ (cooling stop) and
for $340K<T<395K$ the evolution of $\epsilon_m$ is quite different
from $\epsilon_r$. In order to clearly see this effect we divide
$\epsilon_m$ in  the cooling part  $\epsilon_{mc}$ and the heating
part $\epsilon_{mh}$.  In fig.\ref{fig:tempcyc2}(b) we plot the
difference between  $\epsilon_m$ and $\epsilon_r$.
 Filled downwards arrows corresponds to
cooling ($\epsilon_{mc}-\epsilon_{rc}$) and empty upward arrows to
heating ($\epsilon_{mh}-\epsilon_{rh}$). The difference between
the evolutions corresponding to different cooling procedures is
now quite clear. The system keeps the memory of  its previous
aging history when it is reheated from $T_{min}$. The amplitude of
the memory corresponds well to the amplitude of the aging at
$T_{stop}$ but the temperature $T_m$ of the maximum  is shifted a
few degrees above $T_{stop}$. We checked that this temperature
shift is  independent of $T_{stop}$ for temperatures where aging
can be measured in a reasonable time (from $340K$ to $T_g$). This
effect can be seen in fig.\ref{fig:Tstop} where  the difference
between $\epsilon_m$ and $\epsilon_r$ measured for  three
different $T_{stop}$, is plotted  as a function of $T$. We clearly
see that $T_{m}-T_{stop}$ is independent on $T_{stop}$.
Furthermore the amplitude of the downward relaxation at $T_{stop}$
is a decreasing function of $T_{stop}$. It almost disappears for
$T_{stop}<340K$. For this reason  double memory experiments are
more difficult  in PMMA than in SG. However it has to be pointed
out that if two cooling stops are done the system keeps memory of
both of them \cite{elsewhere}.

\begin{figure}
\begin{center}
\centerline{\epsfig{figure=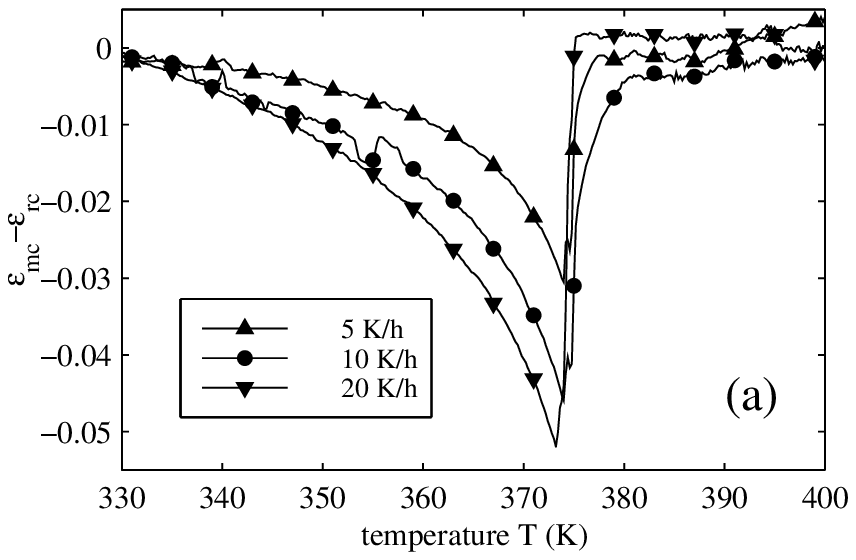} \epsfig{figure=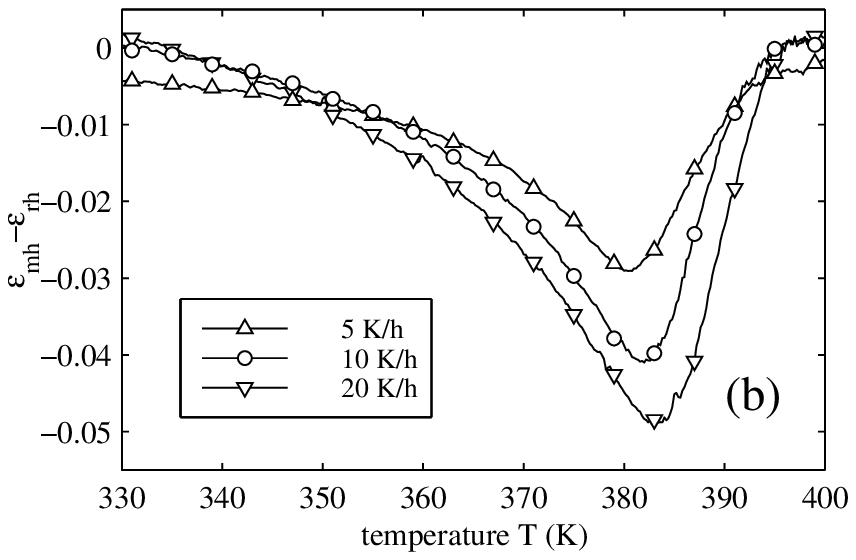}}
\end{center}

\caption{Dependence on the cooling and heating rate. Difference
between  $\epsilon_r$  and $\epsilon_m$ (aging at $T_{stop}=374K$
for $10h$) measured at $\nu=0.1Hz$ for 3 $|R|$. (a) Writing memory
(cooling) : $\epsilon_{mc} - \epsilon_{rc}$ at $5 K/h$
($\blacktriangle$), $10 K/h$ ($\bullet$) and $20 K/h$
($\blacktriangledown$). (b) Reading memory (heating) :
$\epsilon_{mh} - \epsilon_{rh}$ at $5 K/h$ ($\vartriangle$), $10
K/h$ ($\circ$) and $20 K/h$ ($\triangledown$).
 }
\label{fig:rate}
\end{figure}

The memory effect seems to be permanent because it does not depend
on the waiting time at $T_{min}$. Indeed we performed several
experiments in which we waited till $24h$ at $T_{min}$, before
restarting heating, without noticing any change in the heating
cycle. In contrast the amplitude and the position of the memory
effect depend on $R$ and on the measuring frequency. As an example
of rate dependence, at $\nu=0.1Hz$ and waiting time at $T_{stop}$
of $10h$, we plot in fig.\ref{fig:rate} the difference
$\epsilon_m-\epsilon_r$ as a function of $T$ for three different
rates. The faster is the rate the larger is the memory effect and
the farther the temperature of its maximum is shifted above the
aging temperature $T_{stop}$. Finally we checked the dependence of
the memory effect on the measuring frequency.  We find that
 the memory effect becomes larger at the lowest frequency and the
positions of the maxima are at the same temperature.

We can summarize the main results of the low frequency dielectric
measurements on PMMA: (a) The reference curve, obtained at
constant cooling and heating rate is hysteretic.  This hysteresis
is maximum a few degrees above $T_g$. (b) The hysteresis of
$\epsilon_r$ increases with $|R|$.(c) Writing memory : a cooling
stop produces a downward relaxation of $\epsilon_1$. The amplitude
of this downward relaxation depends on $T_{stop}$ and it decreases
for decreasing $T_{stop}$. It almost disappears for
$T_{stop}<330K$. (d) When cooling is resumed $\epsilon_1$ goes
back to the cooling branch of the reference curve. This suggests
that the low temperature state is independent on the cooling
history.  (e) Reading memory : upon reheating $\epsilon_1$ keeps
the memory of the aging history and the cooling stop ({\it
Memory}). The maximum of the memory effect is obtained a few
degrees above $T_{stop}$. (f) The memory effect does not depend on
the waiting time at low temperature but it depends  both on the
cooling and heating  rates \cite{elsewhere}. The memory effect
increases with $|R|$.

 These results can be explained by
a hierarchical free energy landscape, whose barriers growth when
temperature is lowered \cite{Vincent,VincentPRL}. However the
dependence of the memory effect and the hysteresis on $|R|$ and
the independence on the waiting time at $T_{min}$ means that, at
least for PMMA, the free energy landscape has to depend not only
on temperature but also on $|R|$ . The existence of the hysteresis
and temperature shift of the memory effect could also be explained
by a dependence of the landscape on the sign of the rate (and not
only on its magnitude). Many models
\cite{Vincent,Mezard,Fisher,Bray} and numerical simulations
\cite{Marinari,Barrat} do not take into account this dependence
because they consider just a static temperature after a quench. In
contrast points a),b),e) and f) indicate that whole temperature
history is relevant too. Other models based on the idea of domains
growth explain the rate dependence but not the memory effect
\cite{VincentEPL,Bouchaud2}.

Analogies between point a-b) for the hysteresis and point e-f) for
the rate dependence of the memory effect leads to a new
interpretation of hysteresis, which  can be seen as the memory of
aging at a temperature $T_{stop}\approx T_g$. Indeed, in a free
energy landscape model, when cooling the sample just above $T_g$
the system is in its equilibrium phase, that is  in a favorable
configuration at this temperature. If this configuration is not
strongly modified by aging at lower temperatures then,  when
heating back to $T_g$, the system keeps the memory of this
favorable state, just as it does in the memory effect.

It is interesting  to discuss the analogies and the differences
between this experiment and similar ones performed on SG
\cite{VincentPRL,Jonsson1} and  on  OG \cite{Doussineau}. It turns
out that, neglecting the hysteresis of the reference curve of PMMA
and of OG, the behavior of these materials  is quite similar to
that of SG. During the heating period PMMA, SG and OG keep the
memory of their aging history, although the precise way, in which
history is remembered, is material dependent.Furthermore in these
materials the low temperature state is independent on the cooling
history (same response, same aging properties \cite{elsewhere}).
One can estimate the temperature range $\delta T$ where the
material response is different from that of the reference curve
because of the cooling stop. It turns out that the ratio $\delta
T/ T_G $ is roughly the same in PMMA, in  SG and in OG,
specifically $\delta T /T_G \simeq 0.2$.  The important difference
between SG and PMMA is that the amplitude of the downward
relaxation is a function of $T_{stop}$ in PMMA and it is not in
SG.

As a conclusion the "memory"  effect seems to be an  universal
feature of aging whereas  the hysteresis is present in PMMA and in
OG but not in all kinds of spin glasses. It would be interesting
to know if these effects are observed in other polymers and in
supercooled liquids, and if the hysteresis interpretation in terms
of a memory effects hold for other materials. As far as we know no
other results are available at the moment.

We acknowledge useful discussion with J. Kurchan and technical
support by P. Metz and  L. Renaudin. This work has been partially
supported by the R\'egion Rh\^one-Alpes contract ``Programme
Th\'ematique : Vieillissement des mat\'eriaux amorphes''  .

\newpage



\newpage

\begin{thebibliography}{}
\bibitem{comment} In a temperature interval of a few degrees around
 $T_g$ the  PMMA Young modulus changes of several orders of
 magnitude \cite{bookp}. This $T_g$ is consistent with the experimental
evidence  \cite{Bellon} that for $T<T_g$ PMMA presents aging
whereas for $T>T_g$ PMMA relaxes to its equilibrium in less than
$5h$.
\bibitem{elsewhere} More details wil be given elsewhere.
\end{thebibliography}
\end{document}